\documentstyle[aps,epsf]{revtex}

\begin{document}
\title{\begin{flushright} \begin{small}
    hep-th/0106263
\end{small} \end{flushright}
Viscous dissipative effects in isotropic brane cosmology}
\author{Chiang-Mei Chen\thanks{E-mail: cmchen@phys.ntu.edu.tw}}
\address{Department of Physics, National Taiwan University,
Taipei 106, Taiwan}
\author{T. Harko\thanks{E-mail: tcharko@hkusua.hku.hk}}
\address{Department of Physics, The University of Hong Kong,
Pokfulam, Hong Kong}
\author{M. K. Mak\thanks{E-mail: mkmak@vtc.edu.hk}}
\address{Department of Physics, The Hong Kong University of Science
and Technology, Clear Water Bay, Hong Kong}
\date{June 20, 2001}
\maketitle

\begin{abstract}
We consider the dynamics of a viscous cosmological fluid in the
generalized Randall-Sundrum model for an isotropic brane. To
describe the dissipative effects we use the Israel-Hiscock-Stewart
full causal thermodynamic theory. In the limiting case of a stiff
cosmological fluid with pressure equal to the energy density, the
general solution of the field equations can be obtained in an
exact parametric form for a cosmological fluid with constant bulk
viscosity and with a bulk viscosity coefficient proportional to
the square root of the energy density, respectively. The obtained
solutions describe generally non-inflationary brane worlds,
starting from a singular state. During this phase of evolution the
comoving entropy of the Universe is an increasing function of
time, and thus a large amount of entropy is created in the brane
world due to viscous dissipative processes.
\end{abstract}


\section{Introduction}
The idea \cite{RS99a,RS99b} that our four-dimensional Universe
might be a three-brane embedded in a higher dimensional space-time
has recently attracted much attention. According to the
brane-world scenario, the physical fields in our four-dimensional
space-time, which are assumed to arise as fluctuations of branes
in string theories, are confined to the three brane, while gravity
can freely propagate in the bulk space-time, with the
gravitational self-couplings not significantly modified. This
model arisen from the study of a single $3$-brane embedded in five
dimensions, with the $5D$ metric given by $ds^2=e^{-f(y)}
\eta_{\mu\nu} dx^\mu dx^\nu + dy^2$, which can produce a large
hierarchy between the scale of particle physics and gravity due to
the appearance of the warp factor. Even if the fifth dimension is
uncompactified, standard $4D$ gravity is reproduced on the brane.
Hence this model allows the presence of large or even infinite
non-compact extra dimensions and our brane is identified to a
domain wall in a $5$-dimensional anti-de Sitter space-time.

The Randall-Sundrum model was inspired by superstring theory. The
ten-dimensional $E_8\times E_8$ heterotic string theory, which
contains the standard model of elementary particle, could be a
promising candidate for the description of the real Universe. This
theory is connected with an eleven-dimensional theory, M-theory,
compactified on the orbifold $R^{10}\times S^1/Z_2$ \cite{HW96}.
In this model we have two separated ten-dimensional manifolds. For
a recent review of dynamics and geometry of brane Universes see
\cite{Ma01}.

The static Randall-Sundrum solution has been extended to
time-dependent solutions and their cosmological properties have
been extensively studied \cite{KK00} - \cite{Fr01}. In one of the
first cosmological applications of this scenario it was pointed
out that a model with a non-compact fifth dimension is
potentially viable, while the scenario which might solve the
hierarchy problem predicts a contracting Universe, leading to a
variety of cosmological problems \cite{CGKT99}. By adding
cosmological constants to the brane and bulk, the problem of the
correct behavior of the Hubble parameter on the brane has been
solved by Cline, Grojean and Servant \cite{CGS99}. As a result one
also obtains normal expansion during nucleosynthesis, but faster
than normal expansion in the very early Universe. The creation of
a spherically symmetric brane-world in AdS bulk has been
considered, from a quantum cosmological point of view, with the
use of the Wheeler-de Witt equation, by Anchordoqui, Nunez and
Olsen \cite{ANO00}.

The effective gravitational field equations on the brane world, in
which all the matter forces except gravity are confined on the
$3$-brane in a $5$-dimensional space-time with $Z_2$-symmetry have
been obtained, by using an elegant geometric approach, by
Shiromizu, Maeda and Sasaki \cite{SMS00,SSM00}. The correct
signature for gravity is provided by the brane with positive
tension. If the bulk space-time is exactly anti-de Sitter,
generically the matter on the brane is required to be spatially
homogeneous. The electric part of the $5$-dimensional Weyl tensor
$E_{IJ}$ gives the leading order corrections to the conventional
Einstein equations on the brane. The four-dimensional field
equations for the induced metric and scalar field on the
world-volume of a $3$-brane in the five-dimensional bulk with
Einstein gravity plus a self-interacting scalar field have been
derived by Maeda and Wands \cite{MW00}. The effective
four-dimensional Einstein equations include terms due to scalar
fields and gravitational waves in the bulk.

The linearized perturbation equations in the generalized
Randall-Sundrum model have been obtained, by using the covariant
nonlinear dynamical equations for the gravitational and matter
fields on the brane, by Maartens \cite{Ma00}. The nonlocal energy
density determines the tidal acceleration in the off-brane
direction and can oppose singularity formation via the generalized
Raychaudhuri equation. Isotropy of the cosmic microwave background
may no longer guarantee a Friedmann-Robertson-Walker geometry.
Vorticity on the brane decays as in general relativity, but
nonlocal bulk effects can source the gravitomagnetic field, so
that vector perturbations can also be generated in the absence of
vorticity.

A systematic analysis, using dynamical systems techniques, of the
qualitative behavior of the Friedmann-Robertson-Walker (FRW),
Bianchi type I and V cosmological models in the Randall-Sundrum
brane world scenario, with matter on the brane obeying a
barotropic equation of state $p=(\gamma-1)\rho$, has been realized
by Campos and Sopuerta \cite{CS01a,CS01b}. In particular, they
constructed the state spaces for these models and discussed what
new critical points appear, the occurrence of bifurcations and the
dynamics of the anisotropy for both a vanishing and non-vanishing
Weyl tensor in the bulk.

All these investigations of brane cosmological models have been
performed under the simplifying assumption of a perfect
cosmological fluid. But in many cosmological situations an
idealized fluid model of matter is inappropriate, especially in
the case of matter at very high densities and pressures. Such
possible are the relativistic transport of photons, mixtures of
cosmic elementary particles, evolution of cosmic strings due to
their interaction with each other and surrounding matter,
classical description of the (quantum) particle production phase,
interaction between matter and radiation, quark and gluon plasma
viscosity etc. From a physical point of view the inclusion of
dissipative terms in the energy-momentum tensor of the
cosmological fluid seems to be the best motivated generalization
of the matter term of the gravitational field equations.

The first attempts at creating a theory of relativistic fluids
were those of Eckart \cite{Ec40} and Landau and Lifshitz
\cite{LL87}. These theories are now known to be pathological in
several respects. Regardless of the choice of equation of state,
all equilibrium states in these theories are unstable and in
addition signals may be propagated through the fluid at velocities
exceeding the speed of light. These problems arise due to the
first order nature of the theory, that is, it considers only
first-order deviations from the equilibrium leading to parabolic
differential equations, hence to infinite speeds of propagation
for heat flow and viscosity, in contradiction with the principle
of causality. Conventional theory is thus applicable only to
phenomena which are quasi-stationary, i.e. slowly varying on space
and time scales characterized by mean free path and mean collision
time.

A relativistic second-order theory was found by Israel \cite{Is76}
and developed by Israel and Stewart \cite{IsSt76} into what is
called ``transient'' or ``extended'' irreversible thermodynamics.
In this model deviations from equilibrium (bulk stress, heat flow
and shear stress) are treated as independent dynamical variables,
leading to a total of 14 dynamical fluid variables to be
determined. However, Hiscock and Lindblom \cite{HiLi89} and
Hiscock and Salmonson \cite{HiSa91} have shown that most versions
of the causal second order theories omit certain divergence terms.
The truncated causal thermodynamics of bulk viscosity leads to
pathological behavior in the late Universe while the solutions of
the full causal theory are \cite{AnPaRo98}: a) for stable fluid
configurations the dissipative signals propagate causally, b)
unlike in Eckart-type's theories, there is no generic short
wave-length secular instability and c) even for rotating fluids,
the perturbations have a well-posed initial value problem. For
general reviews on causal thermodynamics and its role in
relativity see \cite{Ma95} and \cite{Ma96}.

Causal bulk viscous thermodynamics has been extensively used for
describing the dynamics and evolution of the early Universe or in
an astrophysical context. But due to the complicated character of
the evolution equations, very few exact cosmological solutions of
the gravitational field equations are known in the framework of
the full causal theory. For a homogeneous Universe filled with a
full causal viscous fluid source obeying the relation $\xi \sim
\rho^{1/2}$, exact general solutions of the field equations have
been obtained in \cite{ChJa97,MH98a,MH98b}. In this case the
evolution of the bulk viscous cosmological model can be reduced to
a Painleve-Ince type differential equation, whose invariant form
can be reduced, by means of nonlocal transformations, to a linear
inhomogeneous ordinary second-order differential equation with
constant coefficients \cite{Ch97}. It has also been proposed that
causal bulk viscous thermodynamics can model on a phenomenological
level matter creation in the early Universe \cite{MH98a,MH99a}.
Exact causal viscous cosmologies with $\xi \sim \rho^s$ have been
obtained in \cite{HM99b}.

Because of technical reasons, most investigations of dissipative
causal cosmologies have assumed FRW symmetry (i.e. homogeneity and
isotropy) or small perturbations around it \cite{MaTr97}. The
Einstein field equations for homogeneous models with dissipative
fluids can be decoupled and therefore are reduced to an autonomous
system of first order ordinary differential equations, which can
be analyzed qualitatively \cite{CoHo95,CoHo96}.

It is the purpose of the present paper to investigate the effects
of the bulk viscosity of the cosmological matter fluid on the
dynamics of the brane world. Since the effects of the
extra-dimensions and also the viscous effects are more important
at high matter densities, we restrict our analysis to the extreme
case of a stiff (Zeldovich type) cosmological fluid, with pressure
equal to the energy density. Hence the most important contribution
to the energy density of the matter comes from the quadratic term
in density, and during this period the energy density of matter is
proportional to the Hubble parameter, in opposition to the
standard general relativistic case with energy density
proportional to the square of the Hubble parameter. In this case,
and by assuming that the bulk viscosity coefficient and the
temperature dependence of the cosmic fluid on the energy density
are given by a simple power laws, the field equations can be
solved exactly for several explicit functional forms of the
viscosity coefficient.

The present paper is organized as follows. The field equations on
the brane describing the evolution of a viscous cosmological fluid
are written down in Section II. In Section III we present the
general solution of the field equations for a constant bulk
viscosity and a bulk viscosity coefficient proportional to the
square root of the energy density. The study of the stability of
the equilibrium points of the dynamical system associated to the
evolution of the viscous cosmological fluid is performed in
Section IV. In Section V we discuss and conclude our results.

\section{Dissipative cosmological fluids on the brane}
In the $5D$ space-time the brane-world is located as $Y(X^I)=0$,
where $X^I,\,I=0,1,2,3,4$ are $5$-dimensional coordinates. The
effective action in five dimensions is \cite{MW00}
\begin{equation}
S = \int d^5X \sqrt{-g_5}\left( \frac1{2k_5^2} R_5 - \Lambda_5
\right) + \int_{Y=0} d^4x \sqrt{-g} \left( \frac1{k_5^2} K^\pm -
\lambda + L^{\text{matter}} \right),
\end{equation}
with $k_5^2=8\pi G_5$ the $5$-dimensional gravitational coupling
constant and where $x^\mu,\,\mu =0,1,2,3$ are the induced
$4$-dimensional brane world coordinates. $R_5$ is the $5D$
intrinsic curvature in the bulk and $K^\pm$ is the intrinsic
curvature on either side of the brane.

On the $5$-dimensional space-time (the bulk), with the negative
vacuum energy $\Lambda_5$ and brane energy-momentum as source of
the gravitational field, the Einstein field equations are given by
\begin{equation}
G_{IJ} = k_5^2 T_{IJ}, \qquad T_{IJ} = -\Lambda_5 g_{IJ} +
\delta(Y) \left[ -\lambda g_{IJ} + T_{IJ}^{\text{matter}} \right],
\end{equation}
In this space-time a brane is a fixed point of the $Z_2$ symmetry.
In the following capital Latin indices run in the range $0,...,4$
while Greek indices take the values $0,...,3$.

Assuming a metric of the form $ds^2=(n_I n_J + g_{IJ}) dx^I dx^J$,
with $n_I dx^I = d\chi$ the unit normal to the
$\chi=\text{const.}$ hypersurfaces and $g_{IJ}$ the induced metric
on $\chi=\text{const.}$ hypersurfaces, the effective
four-dimensional gravitational equations on the brane (which are
the consequence of the Gauss-Codazzi equations) take the form
\cite{SMS00,SSM00}:
\begin{equation}  \label{Ein}
G_{\mu\nu} = -\Lambda g_{\mu\nu} + k_4^2 T_{\mu\nu} + k_5^4
S_{\mu\nu} - E_{\mu\nu},
\end{equation}
where $S_{\mu\nu}$ is the local quadratic energy-momentum
correction
\begin{equation}
S_{\mu\nu} = \frac1{12} T T_{\mu\nu} - \frac14 T_{\mu}{}^{\alpha}
T_{\nu\alpha} + \frac1{24} g_{\mu\nu} \left( 3 T^{\alpha\beta}
T_{\alpha\beta} - T^2 \right),
\end{equation}
and $E_{\mu\nu}$ is the nonlocal effect from the bulk free
gravitational filed, transmitted projection of the bulk Weyl
tensor $C_{IAJB}$
\begin{equation}
E_{IJ} = C_{IAJB} n^A n^B, \qquad E_{IJ} \to E_{\mu\nu}
\delta_I^\mu \delta_J^\nu \quad \text{as} \quad \chi \to 0.
\end{equation}

The four-dimensional cosmological constant, $\Lambda$, and the
coupling constant, $k_4$, are given by
\begin{equation}
\Lambda = \frac{k_5^2}2 \left( \Lambda_5 + \frac{k_5^2 \lambda^2}6
\right), \qquad k_4^2 = \frac{k_5^4 \lambda}6.
\end{equation}

The Einstein equation in the bulk, Codazzi equation, also implies
the conservation of the energy momentum tensor of the matter on
the brane,
\begin{equation}
D_\nu T_\mu{}^\nu = 0.  \label{DT}
\end{equation}
Moreover, the contracted Bianchi identities on the brane imply
that the projected Weyl tensor should obey the constraint
\begin{equation}
D_\nu E_\mu{}^\nu = k_5^4 D_\nu S_\mu{}^\nu.  \label{DE}
\end{equation}
Finally, the equations (\ref{Ein}, \ref{DT}) and (\ref{DE}) give
the complete set field equations for the brane gravitational
field.

For any matter fields (scalar field, perfect fluids, kinetic
gases, dissipative fluids etc.) the general form of the brane
energy-momentum tensor can be covariantly given as
\begin{equation}
T_{\mu\nu} = \rho u_\mu u_\nu + p h_{\mu\nu} + \pi_{\mu\nu} + 2
q_{(\mu} u_{\nu)}.  \label{EMT}
\end{equation}
The decomposition is irreducible for any chosen $4$-velocity
$u^\mu$. Here $\rho$ and $p$ are the energy density and isotropic
pressure, and $h_{\mu\nu}=g_{\mu\nu}+u_\mu u_\nu$ projects
orthogonal to $u^\mu$. The energy flux obeys $q_\mu=q_{<\mu>}$,
and the anisotropic stress obeys $\pi_{\mu\nu}=\pi_{<\mu\nu>}$,
where angular brackets denote the projected, symmetric and
tracefree part:
\begin{equation}
V_{<\mu>} = h_\mu{}^\nu V_\nu, \qquad W_{<\mu\nu>} = \left[
h_{(\mu}{}^\alpha h_{\nu)}{}^\beta - \frac13 h^{\alpha\beta}
h_{\mu\nu} \right] W_{\alpha\beta}.
\end{equation}

The symmetric properties of $E_{\mu\nu}$ imply that in general we
can decompose it irreducibly with respect to a chosen $4$-velocity
field $u^\mu$ as
\begin{equation}
E_{\mu\nu} = - k^4 \left[ {\cal U} \left( u_\mu u_\nu + \frac13
h_{\mu\nu} \right) + {\cal P}_{\mu\nu} + 2 {\cal Q}_{(\mu}
u_{\nu)} \right], \label{WT}
\end{equation}
where $k=k_5/k_4$. In Eq. (\ref{WT}) ${\cal U}$ is the effective
nonlocal energy density of the brane arising from the free
gravitational field in the bulk, ${\cal P}_{\mu\nu}$ is the
nonlocal anisotropic stress carying Coulomb, gravito-magnetic and
gravitational wave effects from the bulk, while ${\cal Q}$ is the
effective nonlocal energy flux on the brane.

The effect of the bulk viscosity of the cosmological fluid can be
considered by adding to the usual thermodynamic pressure $p$ the
bulk viscous pressure $\Pi$ and formally substituting the pressure
terms in the energy-momentum tensor by $p_{\text{eff}}=p+\Pi$. The
particle flow vector $N^\mu$ is given by $N^\mu=n u^\mu$, where
$n\geq 0$ is the particle number density.

In the framework of causal thermodynamics and limiting ourselves
to second-order deviations from equilibrium, the entropy flow
vector $S^\mu$ takes the form
\begin{equation}
S^\mu = s N^\mu - \frac{\tau\Pi^2}{2\xi T} u^\mu,
\end{equation}
where $s$ is the entropy per particle, $\tau$ the relaxation time,
$T$ the temperature and $\xi$ is the bulk viscosity coefficient.

We consider that the heat transfer is zero, $q_\mu=0$ in
(\ref{EMT}), and also a vanishing effective nonlocal anisotropic
stress and energy flux, ${\cal P}_{\mu\nu}=0={\cal Q}_\mu$. Then
the matter corrections are given by
\begin{equation}
S_{\mu\nu} = \frac1{12} \rho^2 u_\mu u_\nu + \frac1{12} \rho (\rho
+ 2 p_{\text{eff}}) h_{\mu\nu}.
\end{equation}
The line element of a flat Robertson-Walker metric is given by
\begin{equation}
ds^2 = -dt^2 + a^2(t) \left( dx^2 + dy^2 + dz^2 \right).
\end{equation}
We also assume that the thermodynamic pressure $p$ of the
cosmological fluid obeys a linear barotropic equation of state
$p=(\gamma-1)\rho,\,\gamma=\text{const.}$ and $1\leq\gamma\leq 2$.
Under these assumptions, the field equations and the conservation
equations for the Bianchi type I brane gravitational field take
the form
\begin{eqnarray}
3H^2 &=& \Lambda + k_4^2 \rho + \frac{k_4^2}{2\lambda} \rho^2 +
\frac{6{\cal U}}{k_4^2\lambda}, \label{dH} \\
2\dot H + 3H^2 &=& \Lambda - k_4^2\left[(\gamma-1)\rho + \Pi
\right] - \frac{k_4^2}{2\lambda} \left[ (2\gamma-1)\rho^2 +
2\rho\Pi \right] - \frac{2{\cal U}}{k_4^2\lambda},\label{dVHi}\\
\dot \rho + 3\gamma H \rho &=& -3 H \Pi,  \label{drho} \\
\dot {{\cal U}} + 4H {\cal U} &=& 0,  \label{dU}
\end{eqnarray}
where the Hubble parameter $H$ is defined as $H=\dot a/a$.
$N^\mu{}_{;\mu}=0$ leads to the particle number conservation
equation $\dot n+3Hn=0$.

The causal evolution equation for the bulk viscous pressure $\Pi$
is given by \cite{Ma95}
\begin{equation}
\tau \dot \Pi + \Pi = -3\xi H - \frac12 \tau \Pi \left( 3H +
\frac{\dot \tau}{\tau} - \frac{\dot \xi}{\xi} - \frac{\dot T}{T}
\right).  \label{bulk}
\end{equation}

Eq. (\ref{bulk}) arises as the simplest way (linear in $\Pi$) to
satisfy the $H$-theorem (i.e., for the entropy production to be
non-negative, $S^\mu{}_{;\mu}=\Pi^2/(\xi T)\geq 0$). The
Israel-Stewart theory is derived under the assumption that the
thermodynamic state of the fluid is close to equilibrium, which
means that the non-equilibrium bulk viscous pressure should be
small when compared to the local equilibrium pressure, that is
$|\Pi|<p$.

The growth of the total comoving entropy $\Sigma(t)$ over a proper
time interval $(t_0,t)$ is given by \cite{Ma96}
\begin{equation}
\Sigma(t) - \Sigma(t_0) = - \frac3{k_B} \int_{t_0}^t \frac{\Pi
a^3H}{T} dt.
\end{equation}

An important observational quantity is the deceleration parameter
$q=dH^{-1}/dt-1$. The sign of the deceleration parameter indicates
whether the model inflates or not. The positive sign of $q$
corresponds to ``standard'' decelerating models whereas the
negative sign indicates inflation.

Since the effects of the extra-dimensions are important at very
high densities, when the cosmological fluid behaves like a
Zeldovich fluid with $p=\rho$ ($\gamma=2$), as are also the
dissipative effects, we consider only the physical situation in
which the quadratic term dominates in the energy equation
(\ref{dH}). Therefore during the early period of evolution the
energy density of the Universe is given approximately by $\rho
\approx 3\rho_0 H$, with $3\rho_0=\sqrt{6\lambda/k_4^2}$.

In order to close the system of equations (\ref{dH})-(\ref{bulk})
we have to specify $T$, $\tau$ and $\xi$.

First, following \cite{BNK79} we suppose that the relation $\tau
=\xi/\rho$ holds in order to guarantee that the propagation
velocity of bulk viscous perturbations, i.e. the nonadiabatic
contribution to the speed of sound in a dissipative fluid without
heat flux or shear viscosity does not exceed the speed of light.
An analysis of the relativistic kinetic equation for some simple
cases given by Belinskii and Khalatnikov \cite{BK75}, Belinskii,
Nikomarov and Khalatnikov \cite{BNK79} and Murphy \cite{Mu73} has
shown that in the asymptotic regions of small and large values of
the energy density, the viscosity coefficients can be approximated
by power functions of the energy density with definite
requirements on the exponents of these functions. For small values
of the energy density it is reasonable to consider large
exponents, equal in the extreme case to one. For large $\rho$ the
power of the bulk viscosity coefficient should be considered
smaller (or equal) to $1/2$.

Therefore we assume the following simple phenomenological laws for
the bulk viscosity coefficient, temperature and relaxation time:
\begin{equation}
\xi = \alpha \rho^s = \xi_0 H^s, \qquad T = \beta \rho^r = T_0
H^r, \qquad \tau = \frac{\xi}{\rho} = \frac{H^{s-1}}{\tau_0},
\label{eqstate}
\end{equation}
where $s\geq 0,\, r\geq 0,\, \alpha\geq 0$ and $\beta\geq 0$ are
constants and $\xi_0=\alpha (3\rho_0)^s$ and $\tau_0=\xi_0^{-1}$.

In the context of irreversible thermodynamics, $p, \rho ,T$ and
the number density $n$ are equilibrium magnitudes which are
generally related by equations of state of the form
$\rho=\rho(T,n)$ and $p=p(T,n)$. From the requirement that the
entropy is a state function we obtain the equation
\begin{equation}
\left( \frac{\partial \rho}{\partial n} \right)_T =
\frac{\rho+p}{n} - \frac{T}{n} \left( \frac{\partial p}{\partial
T} \right)_n,
\end{equation}
which imposes the constraint $r=(\gamma-1)/\gamma$. Hence for a
Zeldovich fluid we have $r=1/2$.

With these assumptions the bulk viscous pressure $\Pi$ can be
obtained from Eq. (\ref{drho}) in the form
\begin{equation}
\Pi = -\rho_0 \left( \frac{\dot H}{H} + 6H \right).
\end{equation}

The bulk viscous evolution equation (\ref{bulk}) can be written as
\begin{equation}
\dot \Pi + \frac1{\tau} \Pi = -3\rho H - \frac12 \Pi \left( 3H -
\frac32 \frac{\dot \rho}{\rho} \right),
\end{equation}
and for a stiff cosmological fluid on the brane takes the form
\begin{equation}
\frac{\ddot H}{H} - \frac74 \frac{\dot H^2}{H^2} + \left( 3 +
\tau_0 H^{-s} \right) \dot H + 6 \tau_0 H^{2-s} = 0.  \label{eveq}
\end{equation}

By means of the substitution $H=y^{-4/3}$, Eq. (\ref{eveq}) takes
the form
\begin{equation}
\ddot y + \left( 3 + \tau_0 y^{4s/3} \right) y^{-4/3} \dot y -
\frac92 \tau_0 y^{(4s-5)/3}=0.  \label{eveq1}
\end{equation}

By taking $u=\dot y$ and denoting $v=1/u$, Eq. (\ref{eveq1}) can
be transformed to a second type Abel nonlinear first order
differential equation:
\begin{equation}
\frac{dv}{dy} - \left( 3 + \tau_0 y^{4s/3} \right) y^{-4/3} v^2 +
\frac92 \tau_0 y^{(4s-5)/3} v^3 = 0.  \label{abel}
\end{equation}

By introducing a new variable $\eta=3y^{-1/3}$, Eq. (\ref{abel})
becomes
\begin{equation}
\frac{dv}{d\eta} + \left( 3 + 3^{4s} \tau_0 \eta^{-4s} \right) v^2
- \frac{3^{4s+1}}2 \tau_0 \eta^{1-4s} v^3 = 0.  \label{abel1}
\end{equation}

\section{Brane evolution of dissipative stiff cosmological fluids}
In the previous Section we have formulated the basic equations
describing the dynamics of a dissipative stiff cosmological fluid
on the brane. We have considered only the extreme case of very
high densities, when the main contribution to the energy of the
matter is given by the quadratic term in the energy-momentum
tensor, due to the form of the Gauss-Codazzi equations, and which
leads to major changes in the dynamics of the Universe. In this
case the basic equation describing the evolution of the Universe
can be reduced to an Abel type equation (\ref{abel1}).

It is the purpose of the present Section to consider some exact
classes of solutions of Eq. (\ref{abel1}), corresponding to some
particular values of the constant $s$.

As a first case we assume that the bulk viscosity coefficient
$\xi$ is a constant, $\xi=\xi_0=\text{const.}$, corresponding to
the choice $s=0$ in the equation of state of the bulk viscosity
coefficient. For $s=0$ the temperature and the relaxation time are
functions of density, according to the equations of state
(\ref{eqstate}).

For $s=0$ the evolution equation (\ref{abel1}) of the bulk viscous
pressure takes the form
\begin{equation}
\frac{dv}{d\eta} + \left( 3 + \tau_0 \right) v^2 - \frac32 \tau_0
\eta v^3 = 0.  \label{eqs}
\end{equation}

By introducing a new variable $\eta'=(3+\tau_0)\eta$ and denoting
$b=3\tau_0/2(3+\tau_0)^2$, Eq. (\ref{eqs}) takes the form
\begin{equation}
\frac{dv}{d\eta'} + v^2 - b \eta' v^3 = 0. \label{eqs1}
\end{equation}

By taking $v=w/\eta'$, Eq. (\ref{eqs1}) is transformed into
\begin{equation}
\frac{dw}{d\eta'} = \frac{w}{\eta'}(bw^2-w+1).
\end{equation}

Hence the general solution of Eq. (\ref{eqs1}) is given by
\begin{equation}
\eta' = C \frac{w}{\sqrt{bw^2-w+1}}e^{f(w)},
\end{equation}
where $C>0$ is an arbitrary constant of integration,
\begin{equation}
f(w) = \frac12 \int \frac{dw}{bw^2-w+1},
\end{equation}
and
\begin{eqnarray}
f(w) &=& \frac1{2\sqrt{\Delta}} \ln \left(
\frac{2bw-1-\sqrt{\Delta}}{2bw-1+\sqrt{\Delta}}\right), \quad
\text{if} \quad b<\frac14, \\
f(w) &=& - \frac2{w-2}, \quad \text{if} \quad b=\frac14, \\
f(w) &=&\frac1{\sqrt{-\Delta}} \arctan \left(
\frac{2bw-1}{\sqrt{-\Delta}} \right), \quad \text{if} \quad
b>\frac14,
\end{eqnarray}
where we denoted $\Delta=1-4b$.

Therefore the general solution of the field equations can be
expressed in the following exact parametric form, with $\theta
=1/w$ taken as parameter:
\begin{eqnarray}
t(\theta) - t_0 &=& C_0 \int (\theta^2-\theta+b) e^{-4f(\theta)} d
\theta, \label{const1} \\
H(\theta) &=& H_0 \frac{e^{4f(\theta)}}{(\theta^2-\theta+b)^2}, \\
a(\theta) &=& a_0 \exp \left[ -2 C_0 H_0 f(\theta) \right], \\
q(\theta) &=& 4 \theta/C_0 H_0 - 1, \\
\Pi(\theta) &=& -2 \rho_0 \frac{e^{4f(\theta)}}
{(\theta^2-\theta+b)^2} (3H_0-2\theta/C_0), \\
\Sigma(\theta) &=& \Sigma(\theta_0) +
\frac{6a_0^3\rho_0\sqrt{H_0}}{k_B T_0} \int
\frac{(3C_0H_0-2\theta) \exp[2f(\theta)(1-3C_0H_0)]}
{(\theta^2-\theta+b)^2} d \theta, \\
{\cal U}(\theta) &=& {\cal U}_0 \exp \left[ 8C_0 H_0 f(\theta)
\right], \label{const2}
\end{eqnarray}
where $H_0=[C/3(3+\tau_0)]^4, C_0=3^4(3+\tau_0)^3/C^4$ and $t_0,
a_0$ and ${\cal U}_0$ are constants of integration. In the new
variable $\theta$ the function $f$ is given by $f(\theta)=-2^{-1}
\int (\theta^2-\theta+b)^{-1} d\theta$.

The thermodynamic consistency of the model can be studied from the
ratio of the bulk viscous and thermodynamic pressure, which is
given by
\begin{equation}
l = \left| \frac{\Pi}{p} \right| = \frac13 \left| 5-q \right|.
\label{crit}
\end{equation}

The second case we analyze corresponds to the extreme limit of
very high densities when $s=1/2$. Then Eq. (\ref{abel1}) takes the
form
\begin{equation}
\frac{dv}{d\eta} + \left( 3 + 9 \tau_0 \eta^{-2} \right) v^2 -
\frac{27}2 \tau_0 \eta^{-1} v^3 = 0.  \label{abel2}
\end{equation}

Introducing two new functions $A(\eta)=-2\eta^2/9\tau_0+2/3$ and
$B(\eta)=-2\eta/27\tau_0$ allows to rewrite Eq. (\ref{abel2}) in
the general form
\begin{equation}
\frac{dv}{d\eta} = -\frac{v^3}{B(\eta)} - \left[ \frac{d}{d\eta}
\frac{A(\eta)}{B(\eta)} \right] v^2.  \label{abel3}
\end{equation}

By introducing a new variable
\begin{equation}
\sigma = \frac1{v} - \frac{A(\eta)}{B(\eta)},
\end{equation}
Eq. (\ref{abel3}) can be written in the general form
\begin{equation}
\frac{d\eta}{d\sigma} = B(\eta) \sigma + A(\eta),
\end{equation}
or, equivalently,
\begin{equation}
\frac{d\eta}{d\sigma} = -\frac2{9\tau_0} \eta^2 - \frac2{27\tau_0}
\sigma \eta + \frac23.  \label{ric1}
\end{equation}

Hence we have transformed the initial Abel type equation into a
Riccati equation. A particular solution of Eq. (\ref{ric1}) is
given by
\begin{equation}
\eta = 9 \tau_0 \sigma \Delta(\sigma),
\end{equation}
and therefore the general solution of Eq. (\ref{ric1}) is
\begin{equation}
\eta(\sigma) = 9 \tau_0 \sigma \Delta(\sigma) +
\frac{\Delta^2(\sigma) e^{-\sigma^2/27\tau_0}}{C_1 +
\frac2{9\tau_0} \int \Delta^2(\sigma) e^{-\sigma^2/27\tau_0}
d\sigma},
\end{equation}
where $\Delta(\sigma)=(\sigma^2+27\tau_0/2)^{-1}$.

Hence we obtain the general solution of the gravitational field
equations on the brane for a Zeldovich causal bulk viscous fluid,
with bulk viscosity coefficient proportional to the square root of
the density, in the following exact parametric form, with $\sigma$
taken as parameter:
\begin{eqnarray}
t(\sigma)-t_0 &=& 6 \tau_0^{-1} \int \eta^{-3}(\sigma)d\sigma, \\
H(\sigma) &=& \frac1{81} \eta^4(\sigma), \\
a(\sigma) &=& a_0 \exp \left[ \int \frac{2 \eta(\sigma)}{27\tau_0}
d \sigma \right], \\
q(\sigma) &=& \frac{4\sigma}{\eta(\sigma)} -
\frac{36\tau_0}{\eta^2(\sigma)} + 11, \\
\Pi(\sigma) &=& \frac{2\rho_0}{81} \eta^2(\sigma) [3\eta^2(\sigma)
+ 2 \sigma \eta(\sigma) - 18 \tau_0], \\
\Sigma(\sigma) &=& \Sigma(\sigma_0) - \frac{4\rho_0 a_0^3} {81
\tau_0 T_0 k_B} \int \eta(\sigma) [3 \eta^2(\sigma) + 2 \sigma
\eta(\sigma) - 18 \tau_0] \exp\left[ \int
\frac{2\eta(\sigma)}{9\tau_0} d \sigma \right] d \sigma, \\
{\cal U}(\sigma) &=& {\cal U}_0 \exp \left[ -\frac8{27\tau_0} \int
\eta(\sigma) d \sigma \right].
\end{eqnarray}

\section{Stability analysis of the equilibrium points of the viscous
cosmological fluid}
The general evolution equation of the bulk viscous cosmological
fluid on the brane is given by Eq. (\ref{eveq}). From mathematical
point of view it is a second order non-linear differential
equation of the form $\ddot H + R(H,\dot H)=0$, with $R(H,\dot H)
= -7 \dot H^2/4 H + (3H + \tau_0 H^{1-s}) \dot H + 6\tau_0
H^{3-s}$. Therefore $R(0,0)=\lim_{H,X\to 0} R(H,X)=0$ and $R(H,0)
\neq 0$ for $H \neq 0$.

In order to study the stability of the equilibrium points of the
evolution equation of the viscous cosmological fluid on the brane,
Eq. (\ref{eveq}), we shall rewrite it in the form of an autonomous
dynamical system, by introducing a new variable $X=\dot H$:
\begin{eqnarray}
\frac{d H}{d t} &=& X, \\
\frac{d X}{d t} &=& \frac{7 X^2}{4H} - (3H + \tau_0 H^{1-s}) X - 6
\tau_0 H^{3-s}.
\end{eqnarray}

The critical points of this dynamical system are given by $H=X=0$.
They corresponds to a Minkowskian space-time ($a=\text{const.}=1$)
and to a de Sitter inflationary phase, with $a=\exp(H_0 t),
H_0=\text{const.}$. The system has no other critical points
besides the origin.

The Lyapunov function $V(H,X)$ associated to this system can be
chosen \cite{Wa98} as $V(H,X)=X^2/2+\int_0^H R(s,0) d s$ and is
given by
\begin{equation}
V(H,X) = \frac12 X^2 + \frac{6\tau_0}{4-s} H^{4-s}.  \label{lyap}
\end{equation}
The Lyapunov function (\ref{lyap}) has the properties $V(0,0)=0$
and $dV/dt=7X^3/4H-(3H+\tau_0H^{1-s})X^2$. According to the
standard theory of this type of differential equations
\cite{Wa98} the equilibrium state $(H=0,X=0)$ is stable if the
conditions
\begin{equation}
H R(H,0) = 6 \tau_0 H^{4-s} > 0, \quad \text{for} \quad H \neq 0,
\end{equation}
and
\begin{equation}
\frac{\partial R(H,X)}{\partial X} = -\frac{7X}{2H} + 3 H + \tau_0
H^{1-s} = H \left[\frac72 (q+1)+3+\tau_0 H^{-s} \right] \geq 0,
\end{equation}
holds, where the deceleration parameter $q=-X/H^2-1$. Moreover, if
the condition $\partial R(H,X)/\partial X>0$ is satisfied for
$HX\neq 0$, the equilibrium state is asymptotically stable
\cite{Wa98}. The equilibrium state is unstable if $\partial
R(H,X)/\partial X<0$ for $HX\neq 0$.

The stability criteria of the critical point can be formulated in
terms of some conditions imposed on the deceleration parameter. In
the limit of large time, $H\to 0$ and the term $H^{-s}>0$
dominates in the expression of $\partial R(H,X)/\partial X$,
making it obviously non-negative. In the small time limit,
$H\to\infty $ and the condition of the stability of the critical
point is $7(q+1)/2+3\geq 0$, or $q\geq -13/7$. If $q>-13/7$ the
critical point is also asymptotically stable. On the other hand
for $q<-13/7$ the equilibrium point is unstable.

\section{Discussions and final remarks}
In the present paper we have considered the evolution of a causal
viscous dissipative cosmological fluid in the brane world
scenario. As only source of dissipation we have considered the
bulk viscosity of the matter on the brane. The most important
differences to standard general relativity are expected to occur
in the limit of extremely high densities, when the fluid obeys a
Zeldovich (stiff) equation of state $\rho=p$. In this case the
Friedmann equation is modified due to the presence of the terms
from extra dimensions, quadratic in the energy density, which
dominates the other terms in the energy momentum tensor, leading
to an energy density of the cosmic fluid proportional to the
Hubble parameter.

By assuming the usual equations of state for bulk viscosity,
temperature and relaxation time, the field equations can be solved
exactly for some specific choices of the constant $s$ describing
the bulk viscosity coefficient- energy density functional
relation.

For $s=0$, case corresponding to a constant bulk viscosity
coefficient $\xi=\text{const.}$, the general solution of the field
equations for the viscous fluid on the brane world is given by
Eqs. (\ref{const1})-(\ref{const2}). Since the bulk viscous
pressure $\Pi $ must be negative, $\Pi \leq 0$, it follows that in
order to satisfy this condition the range of the parameter
$\theta$ must be restricted to $\theta \in \lbrack 0,2/3]$.

In the limit of small times, we have $\theta \to 0$ and one
obtains the following equations describing the evolution of the
viscous cosmological fluid on the brane:
\begin{eqnarray}
a &\sim& t^{C_0 H_0/2}, \quad H \sim t^{-2C_0H_0t}, \quad
\rho=p \sim t^{-2C_0H_0t}, \quad q = -1, \nonumber \\
\Pi &\sim& - t^{-1}, \quad {\cal U} \sim t^{-2C_0H_0}, \quad
\Sigma(t) \sim \Sigma(t_0) + t^{3(1-C_0H_0)/2}.
\end{eqnarray}

The variations of the scale factor, Hubble parameter, deceleration
parameter, bulk viscous pressure and comoving entropy of the
constant bulk viscosity coefficient dissipative cosmological fluid
confined on the brane are presented in Figs. 1-5.

\begin{figure}
\epsfxsize=9cm \centerline{\epsffile{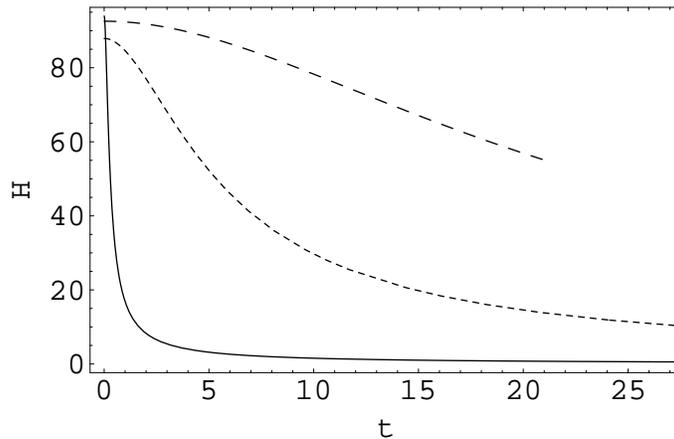}}
\caption{Variation as a function of time of the Hubble parameter
$H$ of the brane Universe with confined dissipative cosmological
fluid with constant coefficient of bulk viscosity for different
values of the parameter $b$: $b=1/6$ (solid curve), $b=1/4$
(dotted curve) and $b=1/2$ (dashed curve). The values of the
constants $H_0$, $C_0$ and $t_0$ are different for each curve.}
\label{FIG1}
\end{figure}

\begin{figure}
\epsfxsize=9cm \centerline{\epsffile{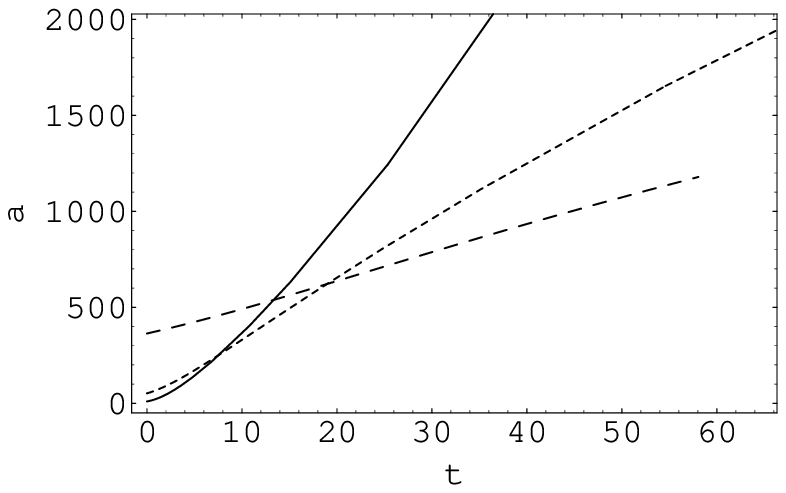}} \caption{Time
evolution of the scale factor $a$ of the brane Universe with
confined dissipative cosmological fluid with constant coefficient
of bulk viscosity for different values of the parameter $b$:
$b=1/6$ (solid curve), $b=1/4$ (dotted curve) and $b=1/2$ (dashed
curve). The values of the constants $a_0$, $C_0$ and $t_0$ are
different for each curve.} \label{FIG2}
\end{figure}

\begin{figure}
\epsfxsize=9cm \centerline{\epsffile{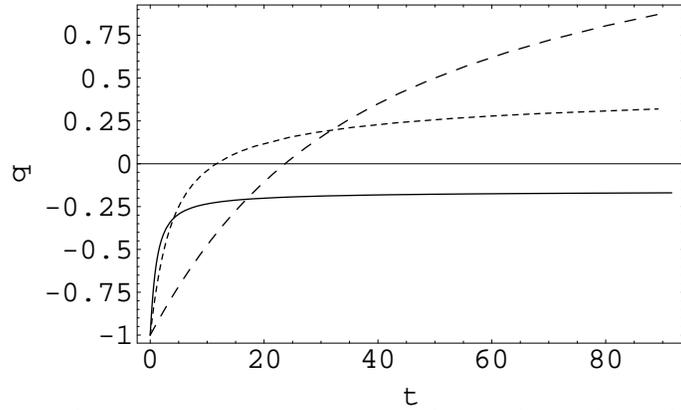}} \caption{Dynamics
of the deceleration parameter $q$ of the brane Universe with
confined dissipative cosmological fluid with constant coefficient
of bulk viscosity for different values of the parameter $b$:
$b=1/6$ (solid curve), $b=1/4$ (dotted curve) and $b=1/2$ (dashed
curve). The value of the constant $t_0$ is different for each
curve and the constants $C_0$ and $H_0$ have been normalized so
that $H_0 C_0=1$.} \label{FIG3}
\end{figure}

\begin{figure}
\epsfxsize=9cm \centerline{\epsffile{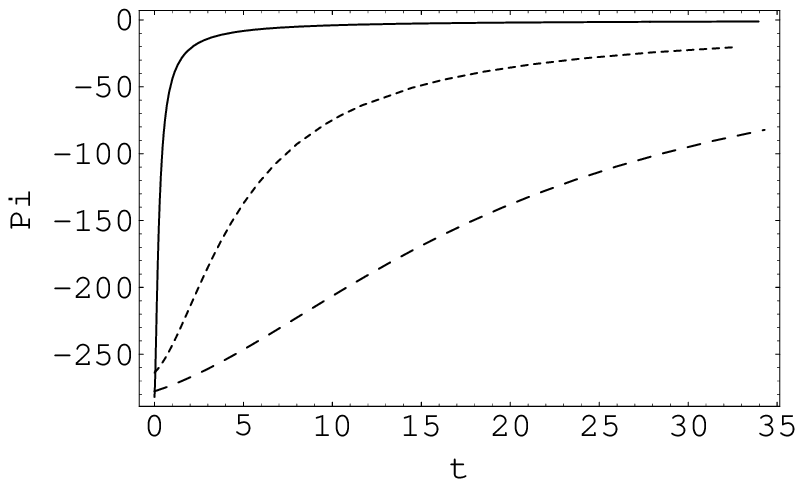}}
\caption{Variation as a function of time of the bulk viscous
pressure $\Pi$ of the brane Universe with confined dissipative
cosmological fluid with constant coefficient of bulk viscosity for
different values of the parameter $b$: $b=1/6$ (solid curve),
$b=1/4$ (dotted curve) and $b=1/2$ (dashed curve). The values of
the constants $H_0$, $C_0$ and $t_0$ are different for each
curve.} \label{FIG4}
\end{figure}

\begin{figure}
\epsfxsize=9cm \centerline{\epsffile{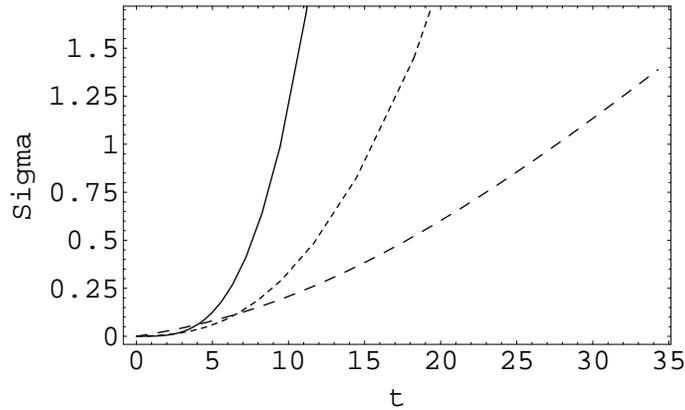}} \caption{Time
evolution of the comoving entropy $\Sigma$ of the brane Universe
with confined dissipative cosmological fluid with constant
coefficient of bulk viscosity for different values of the
parameter $b$: $b=1/6$ (solid curve), $b=1/4$ (dotted curve) and
$b=1/2$ (dashed curve). The values of the constants $H_0$, $C_0$,
$t_0$ and $a_0$ are different for each curve.} \label{FIG5}
\end{figure}

The evolution of the viscous brane Universe starts from a singular
state, with infinite values of the energy density and pressure and
zero scale factor. The initial evolution is inflationary, with a
negative deceleration parameter. But in the large time limit the
dynamics becomes non-inflationary, with the deceleration parameter
$q>0$. Therefore the inclusion of viscous effects during the
period when the quadratic term in energy density (coming from
extra-dimensions) dominates the dynamics of the space time,
provides an effective mechanism for the ``graceful exit'' of the
brane world from the initial inflationary phase to a
non-inflationary era. Due to the dissipative effects the entropy
on the brane is increasing in time and a large amount of entropy
is produced in the early stages of the evolution of the brane
Universe. The nonlocal energy density on the brane, ${\cal U}$, is
a decreasing function of time, so the effects of the gravitational
field on the bulk become rapidly negligeable. But the criterion of
the thermodynamic consistency of the model, $l=|\Pi/p|<1$, is not
generally satisfied in this model, as can be easily seen from Eq.
(\ref{crit}). All the inflationary states clearly contradicts the
condition. On the other hand it is possible to find some
particular of the parameter $b$ leading to thermodynamic
consistency during the noninflationary phase.

The time variations of the Hubble parameter, scale factor,
deceleration parameter, bulk viscous pressure and comoving entropy
for the brane Universe with a dissipative cosmological fluid with
the bulk viscosity coefficient proportional to the square root of
the energy density ($s=1/2$) are represented in Figs. 6-10.

\begin{figure}
\epsfxsize=9cm \centerline{\epsffile{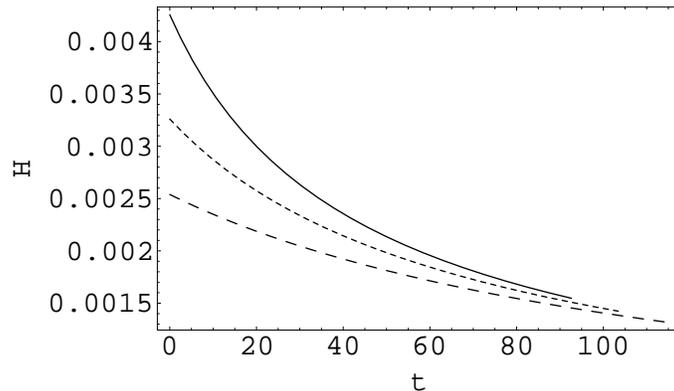}}
\caption{Variation as a function of time of the Hubble parameter
$H$ of the brane Universe with confined dissipative cosmological
fluid with the coefficient of bulk viscosity proportional to the
square root of the energy density ($s=1/2$) for different
numerical values of the integration constant $C_1$: $C_1=0.58$
(solid curve), $C_1=0.62$ (dotted curve) and $C_1=0.66$ (dashed
curve) ($t_0=0$). The constant $\tau_0$ has been normalized so
that $\tau_0=1$.} \label{FIG6}
\end{figure}

\begin{figure}
\epsfxsize=9cm \centerline{\epsffile{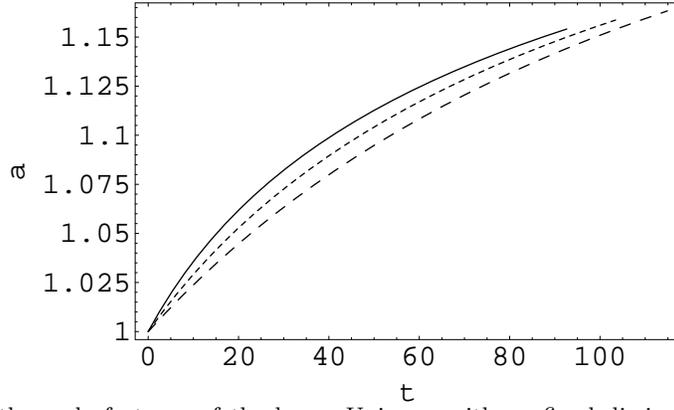}} \caption{Time
evolution of the scale factor $a$ of the brane Universe with
confined dissipative cosmological fluid with the coefficient of
bulk viscosity proportional to the square root of the energy
density ($s=1/2$) for different numerical values of the
integration constant $C_1$: $C_1=0.58$ (solid curve), $C_1=0.62$
(dotted curve) and $C_1=0.66$ (dashed curve) ($t_0=0$). The
constant $\tau_0$ has been normalized so that $\tau_0=1$.}
\label{FIG7}
\end{figure}

\begin{figure}
\epsfxsize=9cm \centerline{\epsffile{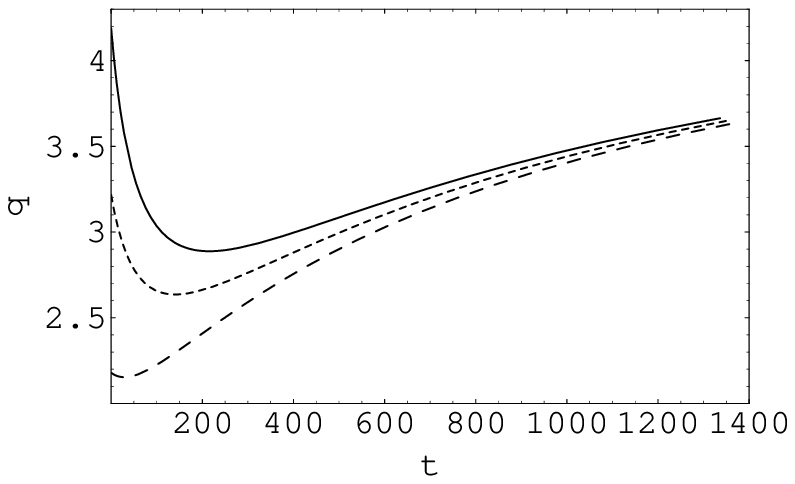}} \caption{Dynamics
of the deceleration parameter $q$ of the brane Universe with
confined dissipative cosmological fluid with the coefficient of
bulk viscosity proportional to the square root of the energy
density ($s=1/2$) for different numerical values of the
integration constant $C_1$: $C_1=0.58$ (solid curve), $C_1=0.62$
(dotted curve) and $C_1=0.66$ (dashed curve) ($t_0=0$). The
constant $\tau_0$ has been normalized so that $\tau_0=1$. The
deceleration parameter satisfies the condition $q>0$ for all
times.} \label{FIG8}
\end{figure}

\begin{figure}
\epsfxsize=9cm \centerline{\epsffile{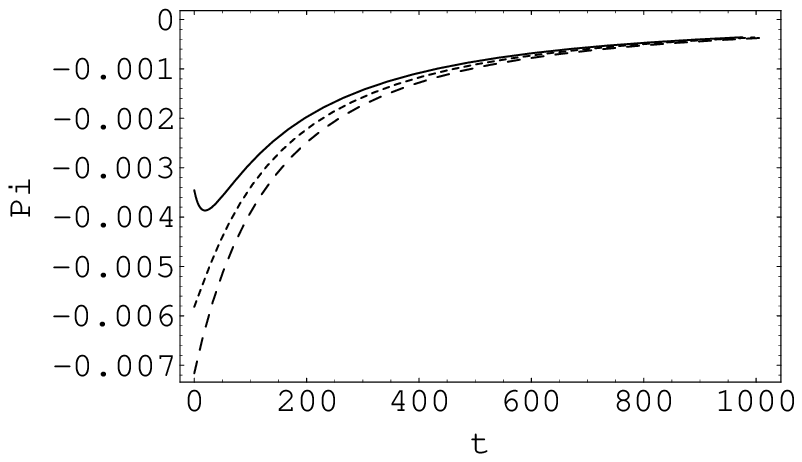}}
\caption{Variation as a function of time of the bulk viscous
pressure $\Pi$ of the brane Universe with confined dissipative
cosmological fluid with the coefficient of bulk viscosity
proportional to the square root of the energy density ($s=1/2$)
for different numerical values of the integration constant $C_1$:
$C_1=0.58$ (solid curve), $C_1=0.62$ (dotted curve) and $C_1=0.66$
(dashed curve) ($t_0=0$). The constant $\tau_0$ has been
normalized so that $\tau_0=1$. As required by the model, the bulk
viscous pressure is negative for all times.} \label{FIG9}
\end{figure}

\begin{figure}
\epsfxsize=9cm \centerline{\epsffile{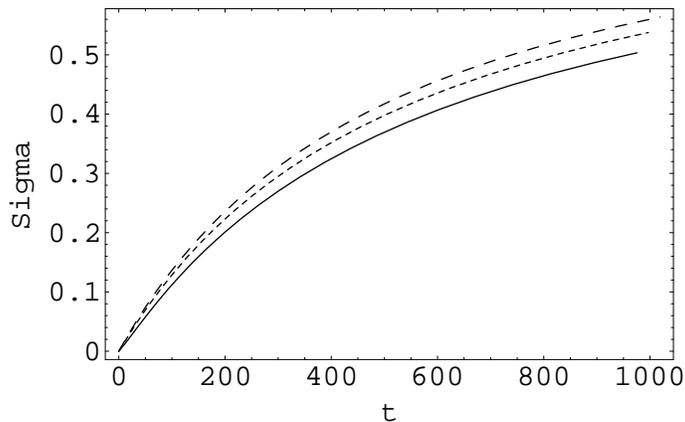}} \caption{Dynamics
of the comoving entropy $\Sigma$ of the brane Universe with
confined dissipative cosmological fluid with the coefficient of
bulk viscosity proportional to the square root of the energy
density ($s=1/2$) for different numerical values of the
integration constant $C_1$: $C_1=0.58$ (solid curve), $C_1=0.62$
(dotted curve) and $C_1=0.66$ (dashed curve) ($t_0=0$). The
constant $\tau_0$ has been normalized so that $\tau_0=1$.}
\label{FIG10}
\end{figure}

The behavior of the Universe depends on the numerical values of
the arbitrary integration constant $C_1$ and of the constant
$\tau_0$. For the chosen numerical values of these parameters the
Universe generally starts from a singular state, with zero and
infinite values of the scale factor and energy density,
respectively. In order the model represents a dissipative fluid,
with negative bulk viscous pressure, the parameter $\sigma$ must
satisfy the condition
$2\sigma<18\tau_0/\eta(\sigma)-3\eta(\sigma)$. In opposition to
the constant bulk viscosity case, the evolution is
non-inflationary for all times. Due to viscous dissipative effects
a large amount of comoving entropy is created on the brane and the
entropy of the Universe is increasing due to viscous dissipation.

Fig. 11 presents the time variation of the ratio $l$ of the bulk
viscous and thermodynamic pressures, respectively. For all times
$|\Pi/p|<1$ and hence in this model the thermodynamical
consistency condition of the smallness of the bulk viscous
pressure is satisfied for all times and for a large class of
admissible values of the integration constant $C_1$ and of
$\tau_0$.

\begin{figure}
\epsfxsize=9cm \centerline{\epsffile{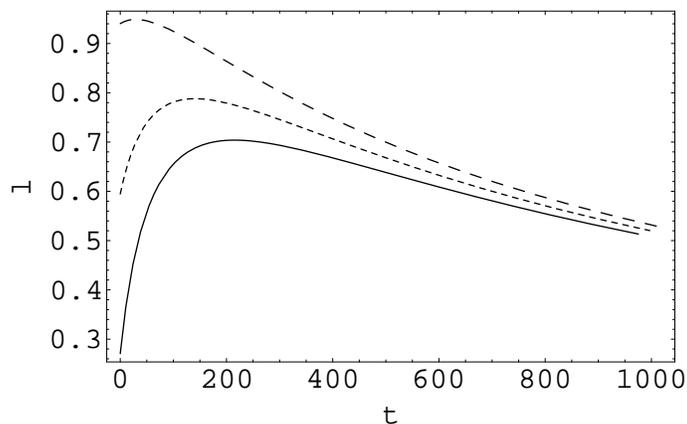}} \caption{Time
variation of the absolute value $l$ of the ratio of the bulk
viscous and thermodynamic pressures, $l=|\Pi/p|$ of the brane
Universe with confined dissipative cosmological fluid with the
coefficient of bulk viscosity proportional to the square root of
the energy density ($s=1/2$) for different numerical values of the
integration constant $C_1$: $C_1=0.58$ (solid curve), $C_1=0.62$
(dotted curve) and $C_1=0.66$ (dashed curve) ($t_0=0$). The
constant $\tau_0$ has been normalized so that $\tau_0=1$. For all
times for the chosen set of parameters the ratio of the pressures
is smaller than $1$.} \label{FIG11}
\end{figure}

The effective nonlocal energy on the brane, ${\cal U}$, is tending
rapidly to zero in the large time limit.

The general exact solution of the gravitational field equations
for a homogeneous flat FRW Universe filled with a causal bulk
viscous fluid with the bulk viscosity coefficient proportional to
the Hubble function, $\xi \sim \rho^{1/2}\sim H$ has been
obtained, in the framework of the standard general relativity
(GR), in \cite{MH98b}. The solution of the field equations can in
this case also be represented in an exact parametric form. There
are major differences between the general evolution of the
dissipative cosmological fluids in the brane and GR models. In GR
the dynamics of the cosmological fluid described by the general
solution is inflationary for all times, with the thermodynamic
consistency condition $|\Pi/p|<1$ violated during the entire
expansionary evolution period. But in the brane model the
evolution is non-inflationary, with the consistency condition
satisfied, at least for a specific range of values of the
parameters $\tau_0$ and $C_1$, which are unknown for a realistic
physical situation (of course for some particular numerical values
of these parameters inflationary initial behavior or increasing
energy density can also be obtained). During the general
relativistic inflationary period the density, temperature, bulk
viscosity coefficient and comoving entropy are rapidly increasing
functions of time. In fact the general solution of the GR field
equations describe a transition between two Minkowskian
space-times connected by an inflationary period. For some
particular values of the parameters one can also obtain general
relativistic non-inflationary solutions \cite{MH98b}.

The consideration of viscous dissipative efects in the brane and
general relativistic models in the extreme limit of very high
densities could be a useful way to differentiate between the two
cosmological scenarios. The different behavior of the energy
density of cosmic matter ($\rho \sim H$ in brane and $\rho \sim
H^2$ in GR, respectively), leads, via the bulk viscous pressure
evolution equation, to differences in the dynamics of the very
early Universe, which perhaps can serve as a tool for testing the
viability of brane model cosmology.

\section*{Acknowledgments}
The work of CMC is supported by the Taiwan CosPA project and, in
part, by the Center of Theoretical Physics at NTU and National
Center for Theoretical Science.


\end{document}